\documentclass[aps,pre,twocolumn]{revtex4-1}
\usepackage{graphicx}
\usepackage{amsmath,amssymb}
\usepackage[utf8]{inputenc}   
\usepackage{color}

\usepackage[T1]{fontenc}
\usepackage{graphicx}  
\usepackage[caption=false]{subfig} 
\usepackage{mathtools}
\usepackage{breqn} 
\usepackage[titletoc]{appendix}

\usepackage[normalem]{ulem} 

\newcommand*{\dd}{\mathrm d}
\newcommand*{\ww}{\mathrm w}
\newcommand*{\hh}{\mathrm h}
\newcommand*{\aaa}{\mathrm a}
\newcommand*{\gGL}{\gamma_{\scriptscriptstyle\rm GL}}
\newcommand*{\gSL}{\gamma_{\scriptscriptstyle\rm SL}}
\newcommand*{\gSG}{\gamma_{\scriptscriptstyle\rm SG}}

\newcommand*{\ttc}{\theta_{\scriptscriptstyle\rm C}}

\newcommand*{\tty}{\theta_{\scriptscriptstyle\rm Y}}

\begin{document}
\title[Paper-Disorder_PRE]
{Geometric and chemical non-uniformity may induce the stability of more than one wetting state in the same hydrophobic surface}

\author{Davi Lazzari, Carolina Brito}
\email{carolina.brito@ufrgs.br}

\date{\today}

\begin{abstract}
It is well established that roughness and chemistry play a crucial role in the wetting properties of a substrate. Yet, few studies have analyzed systematically the effect of the non-uniformity in the distribution of texture and surface tension of substrates on its wetting properties. In this work we investigate this issue theoretically and numerically.
We propose a continuous model that takes into account the total energy required to create interfaces of a droplet in two possible wetting states:  
Cassie-Baxter (CB) with air pockets trapped underneath the droplet; and the other characterized by the homogeneous wetting of the surface,  called the Wenzel (W) state. 
To introduce  geometrical non-regularity we suppose that pillar heights and pillar distances are Gaussian distributed instead of having a constant value. Similarly, we suppose a heterogeneous distribution of Young's angle on the surface to take into account the chemical non-uniformity. This allows to vary the "amount" of disorder by changing the variance of the distribution.
We first solve this model analytically and then we also propose a {\it numerical} version of it, which can be applied to study {\it any type} of disorder.
In both versions, we employ the same physical idea: the energies of both states are minimized to predict the thermodynamic wetting state of the droplet for a given volume and surface texture. We find that the main effect of disorder  is to induce the stability of both wetting states on the same substrate.
In terms of the influence of the disorder on the contact angle of the droplet, we find that it is negligible for the chemical disorder and for  pillar-distance disorder. However,  in the case of pillar-height disorder,  it is observed that the average contact angle of the droplet increases with the amount of disorder.
We end the paper investigating how the region of stability of both wetting states behaves when the droplet volume changes.
\end{abstract}

\maketitle

\section{Introduction}
\label{sec:intro}
Roughness and chemistry of a substrate are key parameters  to understand its  wetting properties \cite{Quere2008}. 
Young understood that when a droplet is placed on an ideal solid, with no texture and a homogeneous chemistry, its contact angle $\tty$ is univocally  determined by minimizing the necessary energies to generate the interfaces of the three involved phases \cite{Young1805}. 
It was later verified that  the apparent contact angle of a droplet  $\ttc$ can be dramatically affected when the substrate is textured or if its chemistry is modified \cite{Wenzel1936, Cassie1944, Borgs1995}. Much advance in controlling the wetting properties of surfaces was possible due to the quantification of the influence of the roughness \cite{Nosonovsky2007, Liu2014} and its chemistry  \cite{Weibel2010}.

Most of the theoretical, numerical \cite{Patankar2003, Marmur2003,  PhysRevE.91.020401, Lundgren2007, Kusumaatmaja2007} and experimental \cite{Tsai2010, Xu2012, McHale2005, Ramos2015, Liu2014} studies approach this problem by varying the roughness via different geometrical parameters  and assuming a {\it regularity} in the distribution of the textures and of the chemistry. 
However, real structures have some degree of disorder in its parameter  \cite{neinhuis1997, Wang2014, ramazanoglu2017} and in fact some simulations and experiments have used wetting dynamics to probe these irregularities  \cite{Pesheva2004, collet2013}. 
Experimental studies have shown that strong spatial disorder have influence on the transition from the CB to W state \cite{boragno2010}, and it has been suggested that the phenomena is related to the negative curvature of the textures \cite{mongeot2010}. The role of disorder has also been studied theoretically for example under the assumption of random distribution of roughness \cite{nayak1973, savva2010, herminghaus2012wetting,Afferrante2015} or irregularities in some types of textures~\cite{collet1997, collet2013}. The results vary depending on the type of disorder: some type of  non-regularities do not influence the wetting properties of the surface, while other types may reduce the droplet contact angle \cite{coninck2013}.
The non-uniformity of the substrate may become relevant when the droplet is small \cite{kim2016} as in experiments where the droplet evaporates and reaches smaller sizes  \cite{Brutin2018, Tsai2010, Chen2012} or to understand the wetting in the case of droplet condensation~\cite{Lv2017,Wen2017,Gao2017}.

In this work we introduce a method that can be used to analyze any type of disordered substrate. We apply this method to study the thermodynamic wetting properties of a pillared surface with three particular types of non-regularities:  a disorder in the distance between pillars, in the height of pillars and in the distribution of $\tty$ on the solid.
When placed on such substrate, the droplet is supposed to be in one of the two wetting states: a Cassie-Baxter (CB)~\cite{Cassie1944} where the droplet resides on the top of the groves, or the Wenzel state (W)~\cite{Wenzel1936}, case where the droplet penetrates the surface.  To describe the wetting properties of these surfaces, we propose two approaches.  i) For each type of substrate, we build a continuous  model that takes into account the global energy necessary to create interfaces between the liquid, air and solid phases when a droplet is placed on a substrate \cite{Sbragaglia2007, Shahraz2012,  Fernandes2015, Silvestrini2017}  and solve it analytically.
 ii) We introduce a  {\it {numerical approach}}   which basically consists in dividing the solid in small areas and numerically look for areas of interfaces  between different phases and then calculate the energy to create them. 
 The advantage of the approach "ii" is its generality: it can be applied for any type of surface, while approach "i" can only be developed for some particular cases of substrates.
 For both approaches, we minimize the energy of both wetting states and the most stable state is the one with smaller energy. 
Among other results, when the substrate presents a disorder of any type considered here, both wetting states, W or CB, may be stable for the same substrate for a certain range of geometrical parameters. 
Concerning the contact angle of the droplet, we find that it can increase when the pillar heights disorder is introduced and it does not change considerably for the other types of disorders.

The paper is organized as follows. Section \ref{sec:continuous_model} explains the continuous model and the process to minimize its equations numerically.  In section \ref{sec:discrete_method} it is introduced the numerical approach
and explained how to obtain the minimum energies of both wetting states.  
 {We present our results for both the analytic and numeric  methods}
in section \ref{sec:results}, which are in very good agreement. We compare the approaches and highlight some of our findings for the particular types of non-regularities considered in this work. We end this section by investigating how the effects of the disorder evolve when the droplet volume changes. In section \ref{sec:conclusion} we draw our conclusions.

\section{The continuous model: analytic approach}
\label{sec:continuous_model}

\begin{figure}[t!]
\includegraphics[width = 0.5\columnwidth]{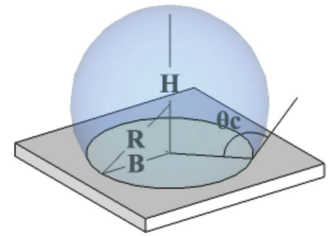}
\caption{{\bf Geometric parameters of the tree-dimensional droplet.~} We  consider that the droplet has a spherical cap with radius $R$, base radius $B$, height $H$ and contact angle $\theta_C$.} 
\label{fig_geometry}
\end{figure}

In this section we develop a model which takes into account all the energies related to the presence of interfaces when a droplet is placed on a textured surface.  The three dimensional droplet considered in this work has  geometric parameters defined in Figure~\ref{fig_geometry}. 
{Throughout this work, we make the following approximations: a) we only consider two wetting states, CB and W, as defined above; mixed states between these two limits are not taken into account. b) We assume a droplet with a spherical cap and c) we disregard pinning of the droplet at the defects of the substrate}.

We first show the equations for a droplet placed on a pillared surface\cite{Fernandes2015}, outlined in Fig.(\ref{fig_sups})-a. This surface is uniform both in terms of geometry -- pillars are distributed regularly -- and in terms of chemistry. We then extend  the model for a droplet placed on the pillared surfaces with three possible types of non-uniformities: i) the distance between pillars not constant, Fig.(\ref{fig_sups})-b, ii) the height of the pillars not constant, Fig.(\ref{fig_sups})-c and iii) the  chemistry of the surface not homogeneous, Fig.(\ref{fig_sups})-d.

\subsection{The continuous model in uniform surfaces}
\label{sec:uniform}

\begin{figure*}[t!]
\centering

\includegraphics[width=.9\textwidth]{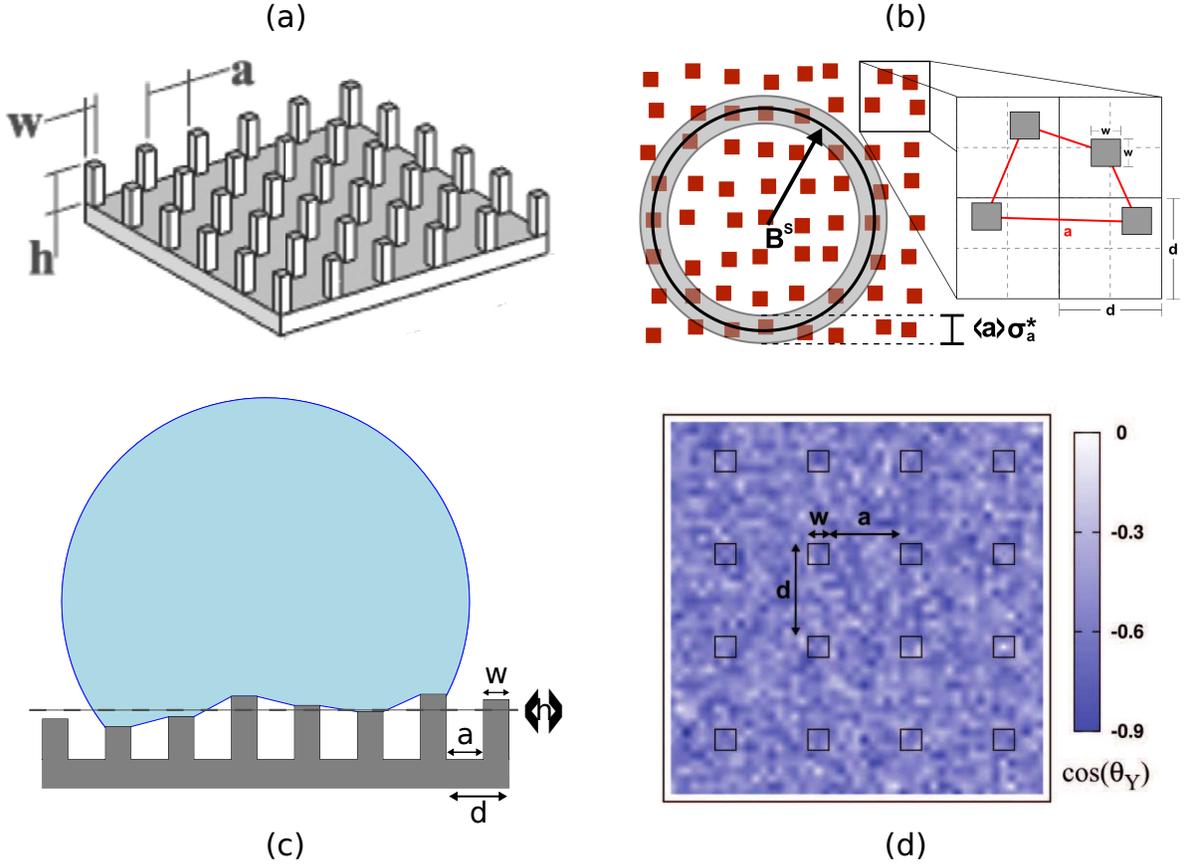}
\caption{{\bf Definition of the parameters for pillared substrates.}  {\bf (a)} The uniform surface is defined by four parameters: the interpillar distance $\aaa$, pillar width $\ww$, pillar height $\hh$ and an Young's contact angle $\tty$.  {\bf (b)} Surface with a Gaussian distribution of the pillar distances characterized by an average distance $\langle \aaa \rangle$ and a normalized standard deviation  $\sigma^*_\aaa$. All other geometric parameters are the same as in (a). For this example,  $\langle \aaa \rangle = 8 \mu m$ and $\sigma^*_\aaa = 0.3$. 
We also show a projection of a droplet with basis $B^S$ and a ring of thickness $\langle \aaa \rangle \sigma^*_\aaa$, which is used to rationalize  how the disorder modifies the energy of the droplet (see text). {\bf (c)} Surface with a Gaussian distribution of pillar heights characterized by an average height $\langle \hh \rangle$, a normalized standard deviation  $\sigma^*_\hh$ and all other geometrical parameters as for the ordered case. In this example, $\ww = 5 \mu m$, $\aaa = 5 \mu m$, $\langle \hh \rangle = 7 \mu m$, $\sigma^*_\hh = 0.3$. It is shown together a droplet deposited on the surface and assuming a CB state to highlight the approximation used in this work: the interface between the liquid and the gas below the droplet is supposed to be linear. {\bf (d)} Example of a surface with chemical non-uniformity, given by a Gaussian distribution with an average  $\langle \cos \tty \rangle = \cos(114^{\circ}) \approx -0.41$ and $\sigma^*_\theta = 0.3$. }
\label{fig_sups}
\end{figure*}

The total energy of each wetting state (W/CB) is given by the sum of all energies involved in creating interfaces between the droplet and  the surface on which it is placed. The difference in  energy of the system with and without the droplet on the surface can be written as $\Delta E^{\rm s}_{\bf Tot} = \Delta E^{\rm s} + E_{\bf g}^{\rm s}$, where the superscript $\rm s$ represents the state  ($\rm s$=W or $\rm s$=CB), $E_{\bf g}$  is the gravitational energy and $\Delta E^{\rm s}$ is the difference in the interfacial energy between every pair formed from  liquid, solid and  gas after the droplet is placed on the surface in  state $\rm s$ and the energy of the surface without the droplet. When the droplet is on the surface,  $E_{\bf g}$  is negligible compared to $\Delta E^{\rm s}~$\cite{Fernandes2015} and for this reason we only take into account the expression for the $\Delta E^{\rm s}$, which for a uniform surface can be written as:
\begin{eqnarray}
\Delta E^{\rm {CB}}_0 &=&  \gGL \left[ {\rm N}^{\rm CB} ~( \underbrace{(\dd^2 - \ww^2)}_{A} -  \ww^2 \cos\theta_Y)  +  S ^{\rm CB} \right], \label{en0_CB} \\
\Delta E^{\rm {W}}_0 &=&  \gGL [ S^{\rm W} - {\rm N}^{\rm W} ( \dd ^2 + 4 \ww \hh)\cos \tty ], \label{en0_W}
\end{eqnarray}
where $\cos\theta_Y = (\gSG - \gSL)/\gGL$ is the Young's equation that describes the wetting behavior of a flat and homogeneous surface. $\gSG$, $\gSL$, $\gGL$ are the solid-gas, solid-liquid and gas-liquid interfacial tension respectively. 
$A$ is the contact area between the liquid and the air trapped under the droplet, $\dd = \ww + \aaa$  and all other geometric parameters are defined in Fig.(\ref{fig_sups})-a. 
The total number of pillars underneath the droplet is 
$N^{\rm s} = \pi (B^{\rm s})^2/ \dd^2$, where $B^{\rm s} = R^{\rm s}\sin(\ttc^{\rm s})$ is the base radius.
The surface area of the droplet cap in contact with air is considered  spherical and is given by $S^{\rm s}=2\pi  {R^{\rm s}}^2 [1-\cos (\ttc^{\rm s})]$. 
When the radius of the droplet is comparable to the roughness of the surface {or the roughness geometry is anisotropic} some deformation in its spherical shape is expected \cite{carmeliet2017, CHEN2005458}, but we will not treat this effect in this work.

Note that the surface tension of the liquid  $\gGL$ multiplies both equations above. It means that this quantity does not influence the thermodynamic stable state of the droplet and therefore we set $\gGL=1$.
The only information about the chemistry of the substrate in the model is contained in  $\tty$.  These considerations will be valid throughout this work.

\subsection{The continuous model in non-uniform surfaces}

We now extend the model for surfaces with non-uniformities (also referred as surfaces with disorder). 
The disorder is introduced by considering that some parameters of the surface, referred as a variable $\xi$, has a Gaussian distribution instead of having a constant value.
The normalized standard deviation, which is defined as the standard deviation divided by the mean of the distribution ($\sigma^* = \sigma / \langle \xi \rangle$), allows us to quantify the disorder for  distributions with different means.
We consider  $\sigma^* \in [0,~0.3]$, where for $\sigma^*= 0.0$  one recovers the case where the surface is uniform (or without disorder) and $\sigma^*= 0.3$ means that $\xi \in [0,~2\langle \xi \rangle]$ \footnote{The probability of finding a value outside of this interval is smaller than $0.1\%$}.

\subsubsection{Geometrical Disorder I: Gaussian distribution of pillar distances ($\aaa$)}
\label{sec:geometric1}

We consider pillared surfaces whose distances "$\aaa$"~between the pillars are given by a Gaussian distribution with mean  $\langle \aaa \rangle$ and a normalized standard deviation $\sigma_{\aaa}^*$. Pillars are not allowed to interpenetrate. An example of this type of substrate is shown in Fig.(\ref{fig_sups})-b, where its geometrical parameters are defined. 

In this same figure we also show a ring of area $A_{ring} = 2 \pi B^{\rm s} \langle \aaa \rangle \sigma_\aaa^*$  and the droplet basis, that is used to model the effect of this type of disorder in the energy of the droplet. The ring is placed on the edge of the droplet and has thickness of $\langle \aaa \rangle \sigma_{\aaa}^*$, which defines the maximum displacement of the pillars.
Note that the pillars inside of the inner disk shown in  Fig.(\ref{fig_sups})-b do not change the energy of the droplet because they cannot leave the droplet basis. However the pillars that are in the outer disk can leave or enter the droplet basis, then altering its energy. 
We estimate that the  number of pillars underneath the droplet can fluctuate as:  
${{\rm N}^{\rm s}_{\aaa} = {\rm N}^{\rm s} \pm \frac{\pi B^{\rm s} \langle \aaa \rangle \sigma^*_\aaa}{\dd^2} }.$
When this result is placed on the energy equations for the ordered surface (Eq.(\ref{en0_CB}) and Eq.(\ref{en0_W})), it is obtained the following equations for the disordered surface:

\begin{dmath}
\Delta E^{\rm {CB}}_{\aaa} = \Delta E^{\rm {CB}}_0 \pm\\ \underbrace{\gGL ~( (\dd^2 - \ww^2) - \ww^2 \cos\tty) \frac{\pi B^{\rm CB} \langle \aaa \rangle \sigma^*_\aaa}{\dd^2}}_{\delta  E^{\rm {CB}}_{\aaa}} , \label{en_a_CB} 
\end{dmath}

\begin{eqnarray}
\Delta E^{\rm {W}}_{\aaa} &=&  \Delta E^{\rm W}_0 \mp \underbrace{\gGL ( \dd ^2 + 4 \ww \hh)\cos \tty \frac{\pi B^{\rm W} \langle \aaa \rangle \sigma^*_\aaa}{\dd^2}}_{\delta E^{\rm {W}}_{\aaa}},. \label{en_a_W} 
\end{eqnarray}

where  geometric parameters of the surface are defined in Fig.(\ref{fig_sups})-b.
We note that the energies can be written as the energies for the ordered surfaces -- $\Delta E^{\rm {CB}}_0$ and $\Delta E^{\rm {W}}_0$ -- plus a {\it dispersion term} around this value which are refereed as $\delta  E^{\rm {CB}}_{\aaa}$ and $\delta  E^{\rm {W}}_{\aaa}$. Clearly, $\sigma_\aaa = 0$ recovers the ordered case.

\subsubsection{Geometrical Disorder II: Gaussian Distribution of pillar heights ($\hh$)}
\label{sec:geometric2}

We now consider pillared substrates such that the value of each pillar height is taken from a Gaussian distribution with mean  $\langle \hh \rangle$ and a normalized standard deviation $\sigma_{\hh}^*$. An example of this type of surface is shown in Fig.(\ref{fig_sups})-c, where we also define the geometric parameters of the surface.

To compute how the  distribution of pillar heights affects the energy difference in the case of the W state, we replace the constant value $\hh$ by  $\hh = \langle \hh \rangle \pm \sigma_{\hh}$ in the Eq.(\ref{en0_W}):

\begin{eqnarray}
\Delta E^{\rm {W}}_{\hh} &=& \Delta E_0^{\rm W} \mp \underbrace{\gGL {\rm N}^{\rm W} \langle \hh \rangle\sigma_\hh^* 4 \ww \cos \tty}_{\delta E^{\rm {W}}_{\hh} }. \label{en_h_W}
\end{eqnarray}

To compute the energy cost for creating interfaces when the droplet is placed on this type of surface and it is in the CB wetting state, we need to compute how the distribution of pillar heights affects the contact area between the gas and the liquid under the droplet, referred in the Eq.(\ref{en0_CB}) by the term $A$. We assume that the interface between the droplet and the gas does not have a meniscus, but instead the interfaces are straight lines as shown in the Fig.(\ref{fig_sups})-c, (actually planes in $3D$). 
This approximation allow us to compute the contact area $A$ between the liquid and the air trapped under the droplet (see Appendix A for details of this computation): 
$A = 2 \ww \sqrt{2 \sigma_\hh ^2 + \aaa^2} + \sqrt{ 3 \sigma_\hh ^4 + 4 \aaa^2 \sigma_\hh ^2 + \aaa^4}.$ 
 We then replace  $A$  in the Eq.(\ref{en0_CB}) to obtain the energy of the state CB in presence of disorder in $\hh$: 
\begin{dmath}
\Delta E^{\rm {CB}}_{\hh} =  \gGL \left[ {\rm N}^{\rm CB} ~( 2 \ww \sqrt{2 \sigma_{\hh}^2 + \aaa^2} + \sqrt{3 \sigma_{\hh}^4 + 4 \aaa^2 \sigma_{\hh}^2 + \aaa ^4} + \ww^2 \cos\theta_Y)  +  S ^{\rm CB} \right]. \label{en_h_CB}
\end{dmath}

In contrast to what happens for the W state,  the energy of the CB state cannot be separated in a part that is the same as in the ordered case plus a dispersion energy term. In this case, the energy only increases when disorder increases and this is a consequence of the fact that $A$ is an increasing function of $\sigma_\hh$, Eq.(\ref{eq_A_append}).

\subsubsection{Chemical non-uniformity: Gaussian Distribution of $\cos \tty$ parameter}
\label{sec:chemical}

As mentioned previously, in our model the dependence of the wetting properties on the chemistry of the surface is contained in the parameter $\tty$. 
Although Young's equation $\cos\theta_Y = (\gSG - \gSL)/\gGL$ relates the interaction between the liquid, the solid and the gas phases, here we assume that the liquid and the gas are  always the same and then changing $\tty$ is an effective way of changing the chemistry of the surface.
We will consider a chemically non-homogeneous surface in such a way that the $\cos \tty $ is Gaussian distributed with a mean value $\langle \cos\tty \rangle$ and a standard deviation $\sigma_\theta$.

Replacing the parameter $\cos \tty$ by the Gaussian distributed one, $\cos \tty = \langle \cos \tty \rangle \pm \sigma_\theta$, in the Eqs. \ref{en0_CB} and \ref{en0_W}, it is obtained the following energy equations:
\begin{eqnarray}
\Delta E^{\rm{CB}}_{\theta} &=& \Delta E^{\rm{CB}}_0 \mp \underbrace{\gGL {\rm N}^{\rm{CB}} \ww^2 \langle \cos\tty \rangle \sigma_{\theta}^*}_{\delta E^{\rm{CB}}_{\theta}} . \label{en_th_CB} \\
\Delta E^{\rm{W}}_{\theta} &=& \Delta E^{\rm{W}}_0 \mp \underbrace{\gGL {\rm N}^{\rm{W}} ( \dd ^2 + 4 \ww \hh) \langle \cos\tty \rangle \sigma_{\theta}^*}_{\delta E^{\rm{W}}_{\theta}} , \label{en_th_W}
\end{eqnarray}

We remind that the geometrical parameters of the surface are defined in the Fig.(\ref{fig_geometry})-d.

\subsection{Energy minimization for the continuous model}
\label{sec:energy_min}

If a droplet with a fixed volume $V_0=4 \pi R_0^3/3~$  is placed on a substrate with a given geometry, the thermodynamic wetting state ${\rm s}$ of the droplet is the one that minimizes its energy  $\Delta E^{\rm s}$. In this section we describe the procedure we employ to compute the minimum energy state, first for the case of an uniform surface \cite{Fernandes2015} and then we extend it for the non-uniform ones.  

{\it Uniform surfaces.~} 
To obtain the minimum energy for the {\rm s=CB} and {\rm s=W} we vary the contact angle between the droplet and the surface $\ttc$  in the interval $(0,\pi]$. An important observation is that the volume of the droplet is a function of its radius $R^{\rm s}$ and contact angle $\ttc$, $V_0=V(R^{\rm s}, \ttc)$. Since we consider a droplet with a fixed volume $V_0$, for each $\ttc$ it is possible to compute the radius $R^{\rm s}$ of the droplet. It is then straightforward to obtain the base radius $B^{\rm s}$, the cap $S^{\rm s}$ and the number of pillars under the droplet ${\rm N}^{\rm s}$,  which in turn defines the energy of the state {\rm s}, $\Delta E^{\rm s}_0$.  In the case of uniform surfaces, we solve Eq.(\ref{en0_CB}) for {\rm s=CB} and Eq.(\ref{en0_W}) for {\rm s=W}.  When we solve these equations numerically for a specified surface and fixed $V_0$, we observe that the curve $\Delta E^{\rm s}_0$ {\it vs} $\ttc$ presents {\it only one minimum state}, called $\Delta E^{\rm s}_{0, \rm minC}$.  If $\Delta E^{\rm W}_{0, \rm minC} < \Delta E^{\rm CB}_{0,\rm minC}$, then W is the thermodynamic stable state. Otherwise CB is the most stable state. We use the subscript {\rm "minC"} to refer to the stable states which are solutions of the continuous model. This is to make a distinction from the solutions of the {numerical approach}
introduced in the next section.

{\it Non-uniform surfaces.~}  Once the minimum wetting states are defined for the uniform case, all geometric parameters of the droplet at the minimum state (contact angle $\ttc$, radius $R^{\rm s}$, base radius $B^{\rm s}$  and spherical cap $S^{\rm s}$) are determined. These data are then used to obtain the dispersion terms for the case of non-uniform surfaces, using the Eq.(\ref{en_a_CB}) and Eq.(\ref{en_a_W}) for the disorder in pillar distances, Eq.(\ref{en_h_W}) and Eq.(\ref{en_h_CB}) for the disorder in pillar heights,  Eq.(\ref{en_th_CB}) and Eq.(\ref{en_th_W}) for  chemical disorder. The states with minimum energy and the dispersion terms found using the numerical minimization of the equations of the continuous model are denoted as $\Delta E^{\rm {\rm s}}_{\aaa,\rm minC}$ $\pm$ $\delta E^{\rm {\rm s}}_{\aaa,\rm minC}$, $\Delta E^{\rm s}_{\hh,\rm minC}$$\pm$ $\delta E^{\rm s}_{\hh,\rm minC}$ and $\Delta E^{\rm s}_{\theta,\rm minC}$ $\pm$ $\delta E^{\rm s}_{\theta,\rm minC}$ for the disorder in pillar distances, pillar heights and chemical disorder respectively.

\section{Numerical Approach}
\label{sec:discrete_method}

In this section we aim to answer to the following question: if a droplet of fixed volume is placed on a non-uniform surface, which is its thermodynamic wetting state? In the previous sections we answered to this question in the case of an ordered substrate and for three particular types of disordered substrates. To do so, we propose a {numerical approach for the continuous model introduced in the previous section, but which can be extended to any type of disordered surface. As previously, the idea is to} take into account the energies in creating the interfaces between different phases (air, liquid and solid) in two possible wetting states, {\rm s=W/CB},  and then minimize the energies to find the global minimum. However, in the case where the substrate has non-regularities of {\it any type}, how can we model the interfacial areas to compute the energy cost in these two wetting states?  To treat this general case, we introduce a numerical approach, as we now explain.

The difference of energy in creating interfaces in the general case where the surface can have any type of disorder can be formally written as:
\begin{eqnarray}
\Delta E^{\rm {W}}_{gen} &=&  \gGL [ S^{\rm W} - A^{\rm W}_{\scriptscriptstyle \rm SL}\cos \tty ], \label{enGEN_W} \\
\Delta E^{\rm {CB}}_{gen} &=&  \gGL \left[  S ^{\rm CB} + A^{\rm CB}_{\scriptscriptstyle \rm GL} - A^{\rm CB}_{\scriptscriptstyle \rm SL} \cos\tty \right], \label{enGEN_CB} 
\end{eqnarray}
where $A^{\rm W}_{\scriptscriptstyle \rm SL}$ and $A^{\rm CB}_{\scriptscriptstyle \rm SL}$ are the interface areas between the liquid and solid phases when the droplet is in the W and CB states respectively and  $(S ^{\rm CB} + A^{\rm CB}_{\scriptscriptstyle \rm GL})$ are the interfacial areas between the liquid and gas for a droplet in the CB state.
To compute the interfacial areas, we cut the substrate in small squares, which we call pixels. Each pixel $i$ has a lateral size $l$ and an area given by $l^2$.   With this procedure, the  interfaces are discrete and can be written as:
\begin{eqnarray}
A^{\rm W}_{\scriptscriptstyle \rm SL} = \sum_i^{n_{\scriptscriptstyle \rm SL}^{\scriptscriptstyle \rm W}} l^2 = l^2 {n_{\scriptscriptstyle \rm SL}^{\scriptscriptstyle \rm W}}
\end{eqnarray}
where $n_{\scriptscriptstyle \rm SL}^{\scriptscriptstyle \rm W}$ is the total number of pixels and  which are in the interface between a solid and a liquid phase for a droplet in the {\rm W} state. Analogously, 
$A^{\rm CB}_{\scriptscriptstyle \rm SL} = l^2 n_{\scriptscriptstyle \rm SL}^{\scriptscriptstyle \rm CB}$ and $A^{\rm CB}_{\scriptscriptstyle \rm GL }= l^2 n_{\scriptscriptstyle \rm GL}^{\scriptscriptstyle \rm CB}$, where $n_{\scriptscriptstyle \rm SL}^{\scriptscriptstyle \rm CB} $ ($n_{\scriptscriptstyle \rm GL}^{\scriptscriptstyle \rm CB}$) is the total number of pixels which are in the interface between a solid and a liquid (gas and a liquid) phase for a droplet in the {\rm CB} state.
For the computation of the energy of the CB wetting state in presence of disorder in the pillar heights, one should take into account the slope of the plane formed by the interface between the gas and the liquid below the droplet.
To take into account the chemical disorder, one can consider that each pixel $i$ has a different value of $\cos\tty$, referred as $\cos\tty^i$. 
To obtain physical energies, we assume $l=1\mu m$. We varied this value, but as soon as $l$ is smaller than the typical sizes of the roughness of the substrate, the results remain unchanged.

\subsection{Energy minimization for the numerical approach}
\label{sec:minimization_discrete}

\begin{figure*}[t!]
\centering

        {\includegraphics[width = 0.95\textwidth]
        {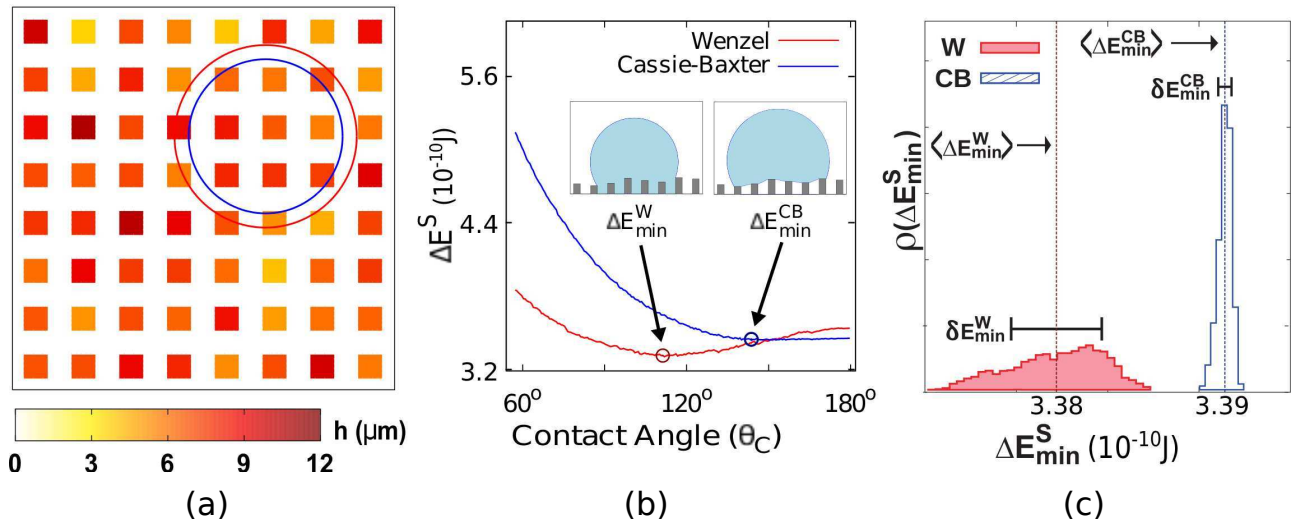}}
\caption{{\bf  Summary of the {numerical approach}}. {\bf  (a)} Top view of a numerically generated surface with disorder in the pillar heights. $\ww = 5 \mu m$, $\aaa = 5 \mu m$, $\langle \hh \rangle = 7 \mu m$, $\sigma^*_\hh = 0.3$ and $\tty=114^{\circ}$. It is also shown the projection of the droplets correspondent to a minimum energy state of a CB state (blue)  W state (red) at a particular position in the surface. {\bf (b)} Numerical solution of $\Delta E^{\rm s}$ {\it vs} $\ttc$ (Eq.(\ref{en_h_W}) and Eq.(\ref{en_h_CB})) for a droplet placed on a position indicated by the circles in (a). Each equation presents only one minimum, $\Delta E^{\rm W}_{\rm min}$ and $\Delta E^{\rm CB}_{\rm min}$. The cross sections correspondent to each wetting state are also shown in the figure. {\bf (c)} Distribution of the $\Delta E^{\rm W}_{\rm min}$ and $\Delta E^{\rm CB}_{\rm min}$, obtained with the procedure of deposing the droplet on the substrate in several positions. Vertical lines represent the mean value of this distributions, $\langle \Delta E_{\rm min}^{\rm s}\rangle$,  and horizontal lines the standard deviation,  $\delta E^{\rm s}_{\rm min}$. }
\label{fig_discrete}
\end{figure*}

Fig.(\ref{fig_discrete})-a shows a top view of the surface used to illustrate the method.
To find the stable wetting state of the droplet when placed on the substrate, we use a similar procedure as previously explained to solve the equations {in the previous section.}
However, since the substrate is non-uniform, its wetting properties can vary in different positions of the surface. To capture this change, we place the droplet in several positions of the substrate and compute numerically its wetting state with minimum energy. For each position,  we adapt the  method used previously 
 as we now explain.

Once a droplet of fixed volume $V_0$ is placed in a particular position of the substrate, we vary  $\ttc$  in the interval $(0,\pi]$. Since  $V_0=V(R^{\rm s}, \ttc)$, for each $\ttc$  we compute the radius $R^{\rm s}$ of the droplet, the base radius $B^{\rm s}$, the cap $S^{\rm s}$,  $n_{\scriptscriptstyle \rm SL}^{\scriptscriptstyle \rm s}$, $n_{\scriptscriptstyle \rm SG}^{\scriptscriptstyle \rm s}$ and $n_{\scriptscriptstyle \rm GL}^{\scriptscriptstyle \rm s}$. We then apply the Eq.(\ref{enGEN_W}) to compute the state s=W and  Eq.(\ref{enGEN_CB}) and compute the energy of the state {\rm s=CB}. 
An example of a numerical solution of these equations as a function of $\ttc$  is shown in Fig.(\ref{fig_discrete})-b,  where we observe that the curve $\Delta E^{\rm s}$ {\it vs} $\ttc$ presents only one minimum state, which we call $\Delta E^{\rm s}_{\rm min}$. In this figure we also show a cross section of these minimum wetting states.

We apply the same procedure for different positions of the surface and this generates one minimal W state and one minimal CB state for {\it each position} of the substrate. After going through the whole surface, we build the distributions of these minima,  shown in the Fig.(\ref{fig_discrete})-c.
We then compute the mean energy and the standard deviation for the state W, $\langle \Delta E^{\rm W}_{\rm min}\rangle$ and $\delta E^{\rm W}_{\rm min}$ , and the same for the state CB, $\langle \Delta E_{\rm min}^{\rm CB}\rangle$ and $\delta E^{\rm CB}_{\rm min}$. The mean energies are represented by the vertical lines in the Fig.(\ref{fig_discrete})-c and the standard deviations are shown by the horizontal lines in the same figure.

The interpretation of the results shown in Fig.(\ref{fig_discrete})-c is that, for this particular substrate, all the minimum energies of the W state are less energetic than the CB minima. Physically it means that the stable wetting state of the droplet would be W in the whole substrate. 

To differ from the minimal states of the continuous model {solved with the analytic approach}, we denote the states with minimum energy and the dispersion terms found using the numerical minimization for discrete model as $\Delta E^{\rm {\rm s}}_{\aaa,\rm min}$ $\pm$ $\delta E^{\rm {\rm s}}_{\aaa,\rm min}$, $\Delta E^{\rm s}_{\hh,\rm min}$$\pm$ $\delta E^{\rm s}_{\hh,\rm min}$ and $\Delta E^{\rm s}_{\theta,\rm min}$ $\pm$ $\delta E^{\rm s}_{\theta,\rm min}$ for the disorder in pillar distances, pillar heights and chemical disorder respectively.

\section{Results and Discussion}
\label{sec:results}

In this section we present and discuss our results. 
The main effect of the disorder is the fact that, depending on the position where the droplet is deposited on the surface, a different wetting state, CB and W, can be the stable one. 
{This phenomenon} is observed for the three types of disorder studied in this work. 
However, the effect of the disorder in the apparent contact angle of the droplet is only relevant for the case of the disorder in pillar heights.  To  quantify these  effects in  substrates with different roughness and disorder, we define what we call 
{\it overlap diagram} as we  explain in the next subsection. 
Since we expect the effect of disorder to be more pronounced in droplets of small size, as the ones reached by the droplet in evaporation experiments \cite{Tsai2010,Xu2012,Ramos2015} or droplet  condensation~\cite{Lv2017,Wen2017,Gao2017}, we  take this limit and  apply the 
overlap diagram
to study the effect of the disorder in the three particular types of disorder for which we have equations of the continuous model {and can solve them analytically}. We compare the theoretical results with the ones obtained using {the numerical approach.} 
It allow us to i) benchmark the {the numerical method introduced in the previous section}  
and ii) discuss the effects of the disorder in the wetting properties in these particular types of non-regularities. We end this section by testing the effect of the geometrical and chemical disorder when the droplet volume increases.

\subsection{{Stability of both wetting states on the same surface}} 
\label{sec:coexistence}

\begin{figure*}[t!]
\centering
        {\includegraphics[width = 0.95\textwidth]
        {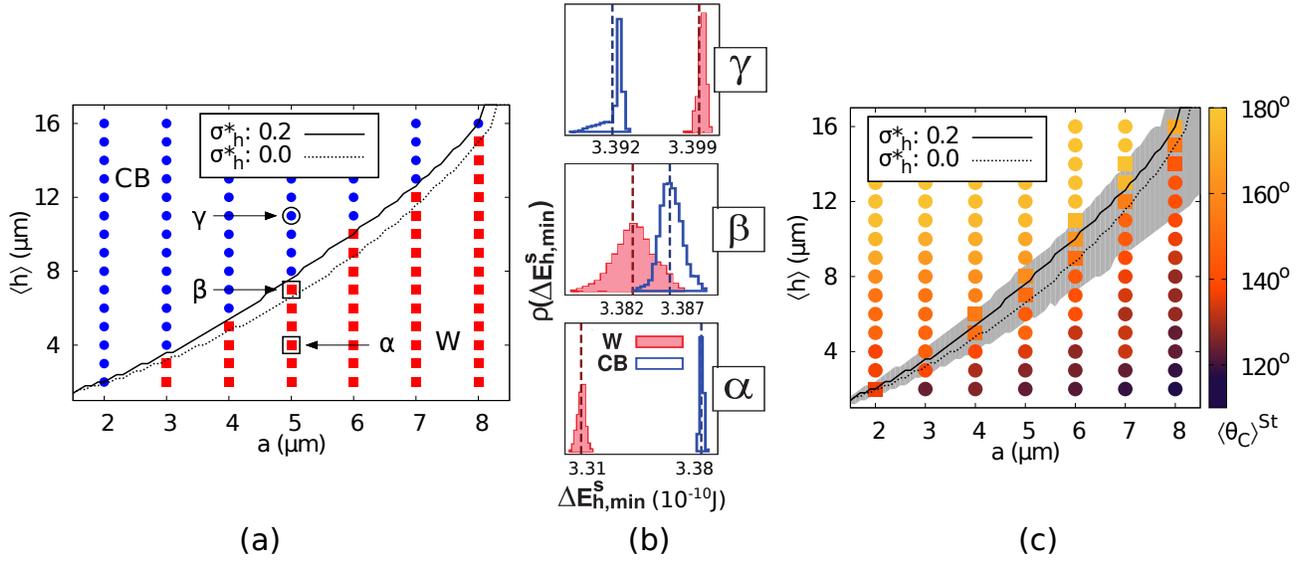}}
        \caption{{\bf (a) Wetting Diagram}. Wetting behavior for  pillared substrate with disorder in pillar heights and geometric parameters varying from $\aaa \in [2,8] \mu m$ and $\langle \hh \rangle \in  [1,16] \mu m$.  Dotted line shows the predicted thermodynamic transition between the CB and W states for the case $\sigma_{\hh}^*=0$ (without disorder) and continuous line for $\sigma_{\hh}^*=0.2$.   Symbols represent the results of the discrete method. Above the transition line $\langle \Delta E_{\rm \hh, min}^{\rm CB}\rangle < \langle \Delta E_{\rm \hh, min}^{\rm W}\rangle$, meaning that the average wetting state of the droplet is CB, while below the transition line the average state is W.  {\bf (b) Distribution of the minimum energy states, $\rho(\Delta E^{\rm s}_{\rm \hh, min})$} for three typical points of the diagram. Blue curve is the distribution of the minimum {\rm CB} states and red for {\rm W} states.  The point $\alpha$ ($\gamma$), below (above) the transition line, shows presents two distributions well separated. Point $\beta$ represents a substrate with a set of geometrical parameters close to the transition line. In this case, the distributions of energy of the two wetting states have an {\it overlap}, indicating that both states could be stable on this substrate. {\bf (c) {Overlap} Diagram.} The overlap exemplified in the point $\beta$ shown in (b) are used to build the  {Overlap} Diagram. Squares represents surfaces for which there is  {overlap} in the solutions of the  {numeric approach} and circles for which there is not. Shaded region is the  {overlap} region identified with the  {analytical solution of the} continuous model. Colors represent the value of the averaged contact angle of the stable wetting state.}
\label{fig_coexistence}
\end{figure*}

To exemplify how to build a {overlap diagram,}
we consider a substrate with disorder in the pillar heights. Fig.(\ref{fig_coexistence})-a presents the diagram of wetting state as a function of the geometric parameters of the surface for fixed values of pillar width and initial droplet radius  ($\ww=5\,\mu m$ and $R_0=100\,\mu m$, respectively)  and varying  $\langle h \rangle \in [1,16]\,\mu m$, and $a \in [2,8]\,\mu m$. Symbols are results of the {numerical method} 
and we now emphasize how they are obtained. 
As explained in the previous section, the droplet is deposited in different positions of the surfaces and, for each position,  the energy of the minimum wetting state {\rm s=W} and {\rm s=CB} are computed. 
Once the whole surface is swept, the mean energy of the minima states are calculated and  denoted as $\langle \Delta E^{\rm W}_{\rm min}\rangle$  and  $\langle \Delta E_{\rm min}^{\rm CB}\rangle$.
To build the diagram of the wetting states, we use these averages to employ the following criterion:  If  $\langle \Delta E_{\rm min}^{\rm CB}\rangle < \langle \Delta E_{\rm min}^{\rm W}\rangle$, then {\rm CB} is the stable state represented by the blue circles. Otherwise {\rm W} is the stable wetting state shown in red  squares in this diagram. 
Lines are results of the continuous model (Eqs. \ref{en_h_W} and \ref{en_h_CB}) taking the equality of the mean energy for both wetting states ($\langle \Delta E^{\rm CB}_{\hh,\rm minC} \rangle = \langle \Delta E^{\rm W}_{\hh, \rm minC} \rangle$). Continuous line are obtained using  $\sigma_{\hh}^*=0.2$ and the dashed line corresponds to the ordered case $\sigma^*_\hh = 0$. 

Fig.(\ref{fig_coexistence})-b shows the distribution of the minimum energies $\Delta E_{\rm \hh, min}^{\rm s}$ obtained for three points of the diagram of Fig.(\ref{fig_coexistence})-a, indicated by $\alpha, \beta, \gamma$.  These distributions correspond to the case with $\sigma^*_\hh = 0.20$. Vertical lines indicate the mean energies $\langle \Delta E_{\hh, \rm min}^{\rm s}\rangle$ for each wetting state.
For the $\alpha$ point we observe that  $\langle \Delta E_{\hh, \rm min}^{\rm W} \rangle < \langle \Delta E_{\hh, \rm min}^{\rm CB}\rangle$  indicating that the average stable state of the substrate is W. Moreover, {\it all} the minimum energies of the W state, $\Delta E_{\rm \hh, min}^{\rm W}$, are smaller than the minimum energies of the CB case,  $\Delta E_{\rm \hh, min}^{\rm CB}$. The physical interpretation of this is that, for any position of the substrate where the droplet is deposited, the stable state is {\rm W}. For the $\gamma$ point the result is the opposite: all the points have  $\Delta E_{\rm  \hh, min}^{\rm CB} <  \Delta E_{\rm \hh, min}^{\rm W}$ which implies that $\langle \Delta E_{\hh, \rm min}^{\rm CB}\rangle < \langle \Delta E_{\hh, \rm min}^{\rm W}\rangle$ and that the most stable wetting state of the droplet placed in any position of this surface would be {\rm CB}. The  most interesting case is the point $\beta$, which lies close to the transition line of the wetting diagram. In this case,  there is an {\it overlap} of the distributions of the different wetting states:  some of the minimum energy CB states ($\Delta E_{\hh, \rm min}^{\rm CB}$) are smaller than some of the minimum energy W states ($\Delta E_{\hh, \rm min}^{\rm W}$). {We have measured the energy difference between both states at each point of the surface, $\Delta=|\Delta E_{\hh, \rm min}^{\rm W} - \Delta E_{\hh, \rm min}^{\rm CB}$|, and have identified that $\Delta$ varies typically between $10^{-17}$J to $10^{-13}$J in the overlap region. Comparing it with the thermal energy at room temperature,  this variation is typically of the other of $100k_BT$ to $10^4K_BT$ where $k_B$ is the Boltzmann constant.}
This implies that, depending on the position where the droplet is deposited on the surface, its stable state can be {\rm CB} or {\rm W} which means that both wetting states can  stable.
We compute this overlap for each point of the wetting diagram to build the 
{\it overlap diagram}
 shown in Fig.(\ref{fig_coexistence})-c. 
Squares represent points of the diagram where the distributions of the minimum energies of the wetting states have an overlap. Note that the points, { which define what we call {\it overlap region},}
lie close to the transition line and its distribution of minima have typically the same behavior as the point $\beta$ in Fig.(\ref{fig_coexistence})-b.  Far from the transition lines, the distributions do not present overlap and are represented by circles. 
 The colors of the symbols indicate the mean contact angle of the stable wetting state, called $\langle \theta_C \rangle ^ {\rm St}$. Shaded region is defined by the { analytical} solutions of the continuous model, which has a good agreement with the 
 {numerical solutions}. 
In the Appendix B we explain mathematically how  the  {overlap}
region is defined.

\subsection{Effect of the disorder for fixed droplet volume}
\begin{figure*}[t!]
\includegraphics[width = 1.25\columnwidth]{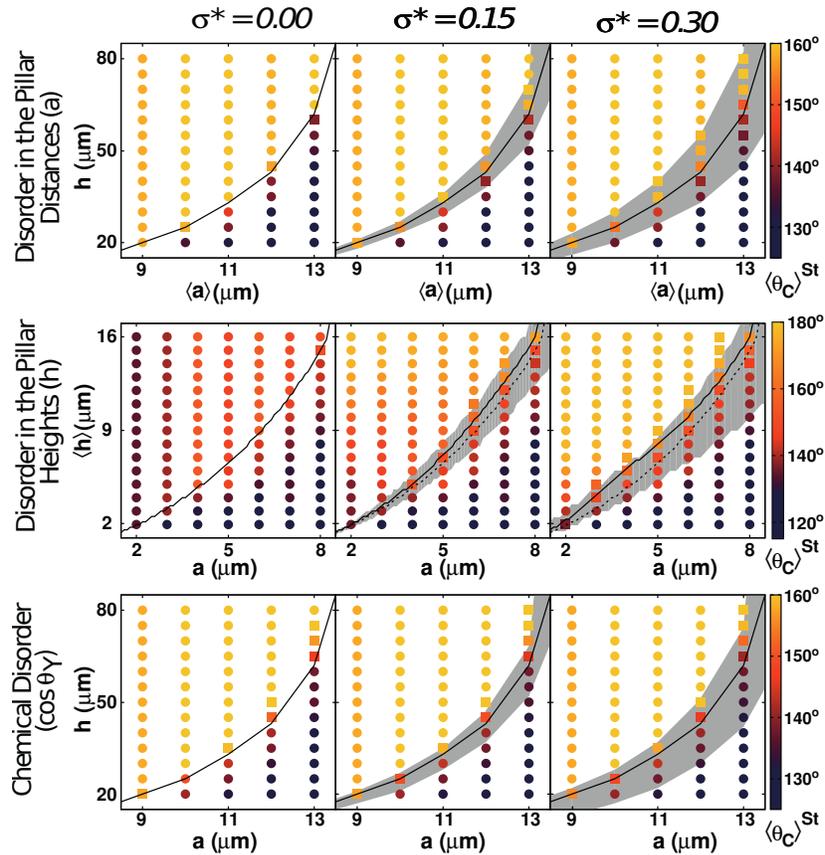}
 \caption{{\bf Influence of the Gaussian disorder on the wetting properties.}   {Overlap} Diagrams for all the three types of disorder (three raws) and for three degrees of disorder (three columns). Shaded regions indicate the   {overlap} region. Transition lines between the states CB (above the line) and W are also shown for the case without disorder (dotted line) and for the disorder case with $\sigma^*$ with a value written on the top of each column (continuous line). Both lines and the shaded region are results of the   {analytic solutions of the}continuous model. Symbols correspond to results of the   {numerical approach}. Squares represent points where there is   {overlap in the distribution} of the wetting states and circles where there is not. Colors indicate the average contact angle of the thermodynamic wetting state and its color bar are shown on the right for each type of disorder. }
\label{fig_summary}
\end{figure*}

In this section we use the  {Overlap} 
Diagrams to explore the effects of different types of disorder on the wetting properties of the substrates. We consider a droplet volume of small size, typically the volume reached when the droplet evaporates \cite{Tsai2010,Xu2012,Ramos2015} or in experiments of droplet  condensation~\cite{Lv2017,Wen2017}.
Figure \ref{fig_summary} shows the  {Overlap} Diagrams as function of the geometric parameters of the surface and fixed values of pillar width and initial droplet radius  ($\ww=5\,\mu m$ and $R_0=100\,\mu m$, respectively). They are built for three different values of normalized standard deviation, which quantifies the amount of disorder.
Based on this figure, we observe:

{\it Transition lines and the  {overlap} regions:~} 
Transition lines are not modified by the disorder in pillar distances and in the case of chemical disorder. This could be anticipated by the equations of energy in these two particular types of disorder because they can be written as the energy of the ordered case and a dispersion term. 
The disorder in pillar heights, however, have a small effect in the transition line: it shifts the line, reducing the CB region.
Reminding that the transition line is defined as $\Delta E^{\rm W}_{\rm minC}=\Delta E^{\rm CB}_{\rm minC}$, we observe that disorder does not influence or have a small influence in the  {\it averaged quantities},  but the relevant effect of the disorder may be observed in the {\it dispersion} around the average, which is responsible for example for the  {\it overlap regions}. In these regions, depending on the place where the droplet is deposited on the substrate, both {\rm CB} or {\rm W} can be the stable states. There regions increase when the disorder increases for the tree types of disorder.
 {In \cite{boragno2010,mongeot2010} they study experimentally the dynamics of the droplet on a surface with randomly distributed pillars (some of them are bent) and show that the spatial disorder can retard the transition from the CB to W state and the wetting dynamics is much more heterogeneous if compared to the ordered case \cite{Sbragaglia2007}. Given that the wetting dynamics is heterogeneous, it would be interesting to investigate if the final state of a  droplet would depend on the place where it is placed on the surface. }

{\it Average apparent contact angle, $\langle \ttc \rangle^{\rm St}$:~} 
It changes very little with the disorder in the pillar distance and with the chemical disorder. We looked at the dispersion of $\ttc$ around $\langle \ttc \rangle^{\rm St}$ and we find that, for points close to the transition lines it has some small deviation, which is at most of $5^\circ$ for the highest value of disorder that we consider here.  
For the case where the disorder is in the pillar heights, it appears and an interesting  effect on the contact angle: when $\sigma_{\hh}^*$ increases, the average contact angle $\langle \ttc \rangle^{\rm St}$ also increases. 
 {The influence of the geometric disorder  on the apparent  contact angle have been studied in \cite{coninck2013, Afferrante2015} and it is found  that its  effect depends on the {\it type of the geometry} of the disorder. In \cite{Afferrante2015} it is considered a randomly rough surfaces and it is shown that the non anisotropy in this type of disorder  is not relevant for the contact angle. In  \cite{coninck2013} they study non-regularities of the substrate with shapes like square protrusions, disks or convex 2D particles and it is found that some of these types of disorder lower the contact angle but  for some cases the contact angle is kept it unchanged.}

{\it Comparison between the results of the  {analytical and numerical approaches}:~} 
There is a good agreement between the results of the  {numerical}
(squares)
and 
 {analytical solutions}
(shaded region) especially above the transition line. Below the transition line, we observe that the 
 {analytical approach overestimate the overlap region.}
Moreover, there is a good agreement for the transition line in both methods.  The agreement is important because the  {analytical approach} 
cannot be applied for any type of disorder, while the  {numerical method} 
is completely general.

\subsection{Effect of disorder as a function of the droplet volume}

\begin{figure}
\includegraphics[width = .95\columnwidth]{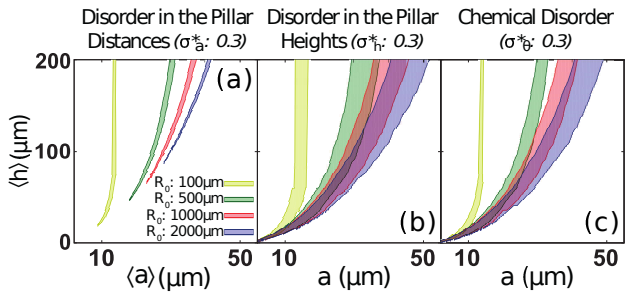}
\caption{{\bf   {Overlap} region for different droplet volume.}   {Overlap} regions for four droplet volumes (indicated by the initial radii $R_0$ in the legend) for the case where there is a disorder in pillar distance {\bf (a)}, disorder in pillar height {\bf (b)} and chemical disorder {\bf (c)}.}
\label{fig_scale}
\end{figure}

In this section we test the effect of the disorder when the droplet volume increases.  We solve  {analytically}  the equations of the continuous model for droplet with varying volumes and measure the  {overlap} 
 region for the case with $\sigma^*=0.3$. 

Figure \ref{fig_scale}-a shows the   {overlap} 
region for the case of the disorder in pillar distance. It evidences the small size of   {overlap} 
region for any value of droplet volume. Due to the fact that for this type of disorder  the energies of the droplet is affected only through the pillars that are on the {\it border of the basis} of the droplet,  it is expected a very small influence in the wetting properties of the substrate and it tends to be less important  when the  droplets gets bigger. 
On the other hand, the  disorder in chemical properties or in pillar heights have influence on the whole basis of the droplet. Then, the effect of these types of disorder in terms of coexistence region is observed for bigger droplet sizes as shown in Fig(\ref{fig_scale})-b,c.
We note, however, that the areas corresponded to the   {overlap}  region are kept roughly constant when the droplet increases. 
We then expect that this effect will be less pronounced in relative terms when the volume gets bigger.

\section{Conclusion}
\label{sec:conclusion}

 In this work we investigate the thermodynamic wetting properties of disordered substrates.
 We first extended a continuous model and a minimization method used for ordered surfaces \cite{Shahraz2012,Fernandes2015,Silvestrini2017} to analyze three  particular cases of disordered surfaces: a pillared substrate  with Gaussian distribution between the distance of the pillars, height of pillars and of $\tty$ (instead of being constant as in the regular case). This choice of disorder allow us to tune the "amount" of disorder by increasing the variance of the distribution. 
We then introduced a   {numerical approach for the same problem.}
The physical idea behind both methods is the same:  to compute the interfacial energies of a droplet placed on a surface in two possible wetting states and then minimize these energies to find the most stable state. 
The advantage of the   {numerical} method is that it can be used to study the wetting properties of any type of surface, including more realistic type of non-regularities ranging from fractal substrates~\cite{Hazlett1990}, or experimental substrate with textures described by KPZ equation~\cite{Almeida2017} or disordered plant surfaces \cite{Wang2014}.

 We find that all types of non-regularities considered in this work have little or no influence on the average quantities as for example average energy of the wetting states. However, disorder does create dispersion which leads to a   {possibility that both wetting states being stable in the same surface}:
  due to the distribution of geometrical parameters or $\tty$, the energies of the wetting states also present a variation and it creates a possibility of finding more than one minimum state in the same substrate.
  One of the interesting aspects of this finding is the association of it
  with the meta-stability encountered in many experimental studies \cite{Quere2008,McHale2005}. It has been reported  that, depending for example on the way that the droplet is deposited on the same surface, its final state can be W or CB. This is usually interpreted as if one of these two states where the stable one and the other were metastable because the droplet would get trapped in a local minimum. Our work offers an alternative interpretation of this, suggesting the possibility of having both states as stable due to the impurities of the substrate.
  

 An important point would be to understand the limit of the effects introduced by the  disorder of the substrate in terms of the droplet volume. 
  This depends on the type of disorder and on the type of phenomena accessed by the experiment.  If one is interested in understanding wetting using evaporation as in some works~\cite{Tsai2010,Ramos2015, Gross2010}, after a certain time the droplet size reaches small  volume and the non-regularities may  play a role. 
 Another example is the recent studies about the droplet  condensation~\cite{Lv2017,Wen2017,Gao2017}. Very recent experimental advances allows to visualize the  initial formation and growing processes of condensed droplets~\cite{Lv2017}. At the initial stages, the individual  droplets have typically a base radius of size  $1 \mu m$, which can be smaller than the typical textures of the surfaces. At these scales, it is expected  impurities of the substrates to play a significant role and as a consequence to influence the dynamics of aggregation of these small droplets to  determine its wetting state. 
 The type of disorder can also play a relevant role if one fabricates a substrate with a distribution of impurities with particular properties. For concreteness, let us suppose a substrate where the distribution of heights follows a power-law distribution like $\rho(\hh) \sim \hh^ \kappa$, with $\kappa < 2$. For this type of distribution, the average of $\hh$ is not defined. Using the same idea as in the model developed in this work for a Gaussian distribution of heights, this would imply that the dispersion of energy of the W state would not have a finite value, leading to an indeterminacy  in the  energy of  this wetting state and perhaps implying the coexistence of more than one wetting state even for droplets of relatively big size.

\acknowledgements
The authors thank Heitor M.C Fernandes for interesting discussions about this work. We also thank the supercomputing of NYU university, where part of the simulations were run, for computer time.

\appendix

\subsection*{Appendix A - Computation of the area under the droplet for CB states with disorder in pillar heights }
\label{append_a}

In this section we show how we compute the interface between the gas and the liquid states for the case where the droplet is placed on a surface with disorder in pillar heights and in the CB state. An example of this situation is shown in the Fig.(\ref{fig_sups})-c.

\begin{figure}[h]
\centering
\includegraphics[width=0.6\linewidth]{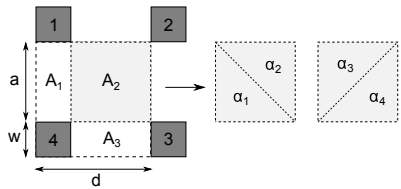}
\caption{Upper view of an "unitary cell" of the plane which is the interface between liquid and gas for the CB state. In the case where pilars have different size, this plane is inclined in respect to the substrate and we divided it in triangles to compute the area.}
\label{fig_append}
\end{figure}

In our model we approximate this interface as a plane and since pillar have different heights, the plane is inclined in respect to the substrate. In Figure \ref{fig_append} it is shown a schema of one "unitary cell" of this plane. It is shown four dark gray pillars with width "$\ww$"~and distance "$\aaa$". The total area $A$ of the unitary cell is the sum of the  areas $A = A_1+A_2+A_3$.  $A_1$ and $A_3$ are formed by the plans connecting pillars 4-1 and 4-3, respectively, and are calculated using the difference between pillar heights as $\Delta \hh_{ij} = \hh_j - \hh_i$, where $j,i$ are neighbor pillars. 
Area $A_2$ is delimited by four points with different heights and can be computed as the average of the four  triangles as indicated in the figure, $A=(\alpha_1+\alpha_2+\alpha_3+\alpha_4)/2$.
Since pillar heights are Gaussian distributed, the difference between the heights of neighboring pillars is independent of the pillar index and is defined as $\Delta \hh$. Then the average value of the area $A$ becomes $A = 2 \langle A_1 \rangle + \langle A_2 \rangle = 2 \langle A_1 \rangle + 2 \alpha$ because $\langle A_1 \rangle = \langle A_3 \rangle$ and $\alpha = \langle \alpha_k \rangle$ for any $k \in \{1,2,3,4\}$.

Considering the geometries of the surface, it is possible to define the areas $\langle A_1 \rangle$ and $\alpha$ depending only on the $\Delta \hh$:
\begin{equation}
A = 2 \ww \sqrt{(\Delta \hh)^2 + \aaa^2} + \sqrt{ \frac{3}{4}(\Delta \hh)^4 + 2\aaa^2 (\Delta \hh)^2 + \aaa^4 }.
\label{eq_A_append}
\end{equation}
Note that for $\Delta \hh=0$, $A=\aaa^2+2\aaa\ww$, recovering the case without disorder in pillar height.

Now, considering the heights as normally distributed random variables, the difference of this values are given by a theorem \cite{GaussianSum}, which predicts that the distribution $\Delta \hh$ have the mean $\langle \Delta \hh \rangle = \langle \hh_i \rangle - \langle \hh_j \rangle = 0$ and the variance $\sigma_{\Delta \hh}^2 = \sigma_{\hh_i}^2 + \sigma_{\hh_j}^2 = 2\sigma_{\hh}^2$, where $i$ and $j$ now represents the position of two neighboring pillars. Taking the difference in the pillar heights in Eq.(\ref{eq_A_append}) given by $\Delta \hh = \langle \Delta \hh \rangle \pm \sigma_{\Delta \hh} = 0 \pm \sqrt{2}\sigma_{\hh}$ we obtain the equation for $A$ for the continuous model for disorder in the pillar heights shown in the main text above Eq.(\ref{en_h_CB}).

\subsection*{Appendix B - Mathematical definition of the   {overlap} region}
\label{append_b}

To define the   {overlap in the case of the analytical solutions}
 of the continuous model, shown by the shaded region in the Fig.(\ref{fig_coexistence})-c we first solve the equations of the energy of the droplet in the two wetting state (the specific equation depends on the type of substrate we are considering)
$\Delta E_{t, \rm minC}^{\rm CB}$ and 
$\Delta E_{t, \rm minC}^{\rm W}$, where $t$ represents the type of disorder ($t$ = $\aaa$, $\hh$ or $\theta$) and also the dispersion terms  $\delta E^{\rm CB}_{t,\rm minC}$ and  $\delta E^{\rm W}_{t,\rm minC}$.

To define the inferior border of the   {overlap} region, we use the following criteria:
\begin{enumerate}
    \item  $\Delta E_{t, \rm minC}^{\rm W}$ $<$ $\Delta E_{t, \rm minC}^{\rm W}$
    \item $\Delta E_{t, \rm minC}^{\rm W}$ + $\delta E^{\rm W}_{t,\rm minC}$ = 
$\Delta E_{t, \rm minC}^{\rm CB}$ - $\delta E^{\rm CB}_{t,\rm minC}$.
\end{enumerate}

For the superior border of the   {overlap} region, the criteria  are:

\begin{enumerate}
    \item $\Delta E_{t, \rm minC}^{\rm CB}$ $<$ $\Delta E_{t, \rm minC}^{\rm W}$
    \item $\Delta E_{t, \rm minC}^{\rm CB}$ + $\delta E^{\rm CB}_{t,\rm minC}$ = 
$\Delta E_{t, \rm minC}^{\rm W}$ - $\delta E^{\rm W}_{t,\rm minC}$.
\end{enumerate}

For the case where t=$\hh$, we cannot take any analytical dispersion in the CB state. Then to calculate the inferior and superior border of the    {overlap} region, we make the same calculation above but we considerer $\delta E^{\rm CB}_{\hh,\rm minC} = 0$. 

 {For the numerical approach} the situation is very similar, but instead of computing the dispersion terms $\delta E^{\rm S}_{t,\rm minC}$ using an equation as in the case of the continuous model, 
we look directly to the minimum energy distribution of each wettability state and analyze the overlapping of these distributions.

\bibliographystyle{apsrev4-1}
\bibliography{droplet}

\end{document}